\documentclass[aps,pra,twocolumn,superscriptaddress]{revtex4}

\usepackage{amsmath,amssymb,amsthm,color,mathrsfs}
\usepackage{amsbsy,mathtools}

\usepackage{hyperref}
\usepackage{graphicx}
\usepackage{ulem}
\usepackage{bm}

\renewcommand{\em}{\it}

\newcommand{\p}{\partial}

\newcommand{\bnabla}{\bm{\nabla}}
\newcommand{\ex}{\bm{\hat{e}}_1}
\newcommand{\ey}{\bm{\hat{e}}_2}
\newcommand{\ez}{\bm{\hat{e}}_3}
\newcommand{\emn}{\epsilon_{\mu\nu}}

\newcommand{\DM}{D}
\newcommand{\dm}{\lambda}
\newcommand{\Anisotropy}{K}
\newcommand{\anisotropy}{\kappa}
\newcommand{\Exchange}{A}

\newcommand{\lex}{\ell_{\rm ex}}
\newcommand{\Energy}{W}

\newcommand{\Ea}{W_{\rm a}}

\newcommand{\Edm}{W_{\rm DM}}
\newcommand{\hext}{h}
\newcommand{\bhext}{\bm{h}}
\newcommand{\xmagn}{X}
\newcommand{\ymagn}{Y}

\newcommand{\tmagn}{\mu}
\newcommand{\mass}{m}

\newcommand{\skyrmion}{q}
\newcommand{\Skyrmion}{Q}

\newcommand{\ldm}{\ell_{\rm D}}

\begin{document}

\title{Skyrmion dynamics in chiral ferromagnets}
\author{Stavros Komineas}
\affiliation{Department of Mathematics and Applied Mathematics, University of Crete, 71003 Heraklion, Crete, Greece}
\author{Nikos Papanicolaou}
\affiliation{Department of Physics, University of Crete, 71003 Heraklion, Crete, Greece}
\begin{abstract}
We study the dynamics of skyrmions in Dzyaloshinskii-Moriya materials with easy-axis anisotropy.
An important link between topology and dynamics is established through the construction of unambiguous
conservation laws obtained earlier in connection with magnetic bubbles and vortices. 
In particular, we study the motion of a topological skyrmion with skyrmion number $\Skyrmion=1$ and a
non-topological skyrmionium with $\Skyrmion=0$ under the influence of an applied field gradient.
The $\Skyrmion=1$ skyrmion undergoes Hall motion perpendicular to the direction of the field gradient
with a drift velocity proportional to the gradient.
In contrast, the non-topological $\Skyrmion=0$ skyrmionium is accelerated in the direction of the field gradient,
thus exhibiting ordinary Newtonian motion.
When the applied field is switched off 
the $\Skyrmion=1$ skyrmion is spontaneously pinned around a fixed guiding center,
whereas the $\Skyrmion=0$ skyrmionium moves with constant velocity $v$.
We give a systematic calculation of a skyrmionium traveling with any
constant velocity $v$ that is smaller than a critical velocity $v_c$.
\end{abstract}


\maketitle

\section{Introduction}
\label{sec:intro}

Magnetization structures in the form of topological solitons have come to the center of research interest 
following the fabrication of various ferromagnetic materials.
Magnetic bubbles were intensively studied in the 60s and 70s for technological applications \cite{MalozemoffSlonczewski},
and theoretically predicted magnetic vortices were experimentally observed as ground states 
in mesoscopic magnetic elements in the 90s \cite{HillebrandsOunadjela2003,BlandMills2006}.
Stable topological solitons had been predicted also in the presence of the Dzyaloshinskii-Moriya (DM) interaction
\cite{BogdanovYablonskii_JETP1989,BogdanovHubert_JMMM1994}.
They were observed in recent years as
isolated structures \cite{RommingHanneken_Science2013,RommingKubetzka_PRL2015} or forming lattices \cite{MuhlbauerBinz_Science2009,YuOnose_Nature2010,YuKanazawa_NatMat2011}

The dynamics of topological magnetic solitons, such as bubbles and vortices, has been long 
recognized to display peculiar features \cite{Thiele_PRL1973}.
They are deflected in a direction almost perpendicular to an applied magnetic field gradient
\cite{MalozemoffSlonczewski} and their dynamics
is seen to be similar to the Hall motion of a charged particle in the presence of a magnetic field.
A theoretical description links the dynamics of topological solitons with their nonzero 
topological number, by example of magnetic bubbles 
\cite{PapanicolaouTomaras_NPB1991,KomineasPapanicolaou_PhysD1996}.

The opportunity now arises to study peculiar soliton dynamics 
for the case of skyrmions in DM materials.
These materials are well suited for such studies for two main reasons.
The DM interaction introduces an intrinsic length scale which defines the size of the skyrmion,
therefore the skyrmion is expected to be robust and remain rigid under external probes.
In the presence of easy-axis anisotropy we have a topological skyrmion with skyrmion number $\Skyrmion=1$  
as well as non-topological $\Skyrmion=0$ solitons ($2\pi$ vortices) \cite{BogdanovHubert_JMMM1999,LeonovRoessler_EPJ2013}
thus allowing to explore theoretical predictions for dramatically different dynamical behaviors.

Skyrmions could be the stable and robust entities that are needed for the technology of
recording and transferring information, currently mainly obtained in magnetic media using domain walls.
Current induced motion is under intensive study and it is a promising technique 
for the manipulation of magnetic information
\cite{FertCros_NatNano2013,SampaioCros_NatNano2013,NagaosaTokura_NatNano2013,IwasakiMochizuki_NatNano2013,ButtnerMoutafis_NatPhys2015,WooLitzius_arXiv2015}.
The different dynamical behaviors which will be described in this paper can help 
skyrmions emerge as particularly attractive entities for applications in the area of  transfer of magnetic information.

Sec.~\ref{sec:model} gives a description of the model, 
Sec.~\ref{sec:skyrmions} discusses static $\Skyrmion=1$ and $\Skyrmion=0$ solutions,
Sec.~\ref{sec:dynamics} contains a review of the theory and gives the main results of the dynamics of
skyrmions and Sec.~\ref{sec:conclusions} contains our concluding remarks.

\section{The model}
\label{sec:model}

We assume a thin film with easy-axis anisotropy perpendicular to the $xy$-plane of the film
and with a DM energy term \cite{BogdanovHubert_JMMM1994}.
If $\bm{M}=\bm{M}(x,y)$ is the magnetization vector the energy functional reads
\begin{equation}  \label{eq:energy0}
\begin{split}
\Energy & = \frac{\Exchange}{M_s^2} \int  \p_\mu \bm{M}\cdot \p_\mu \bm{M}\, dx dy + \frac{\Anisotropy}{M_s^2} \int (M_1^2+M_2^2)\, dx dy  \\
             & + \frac{D}{M_s^2} \int \left[ (M_1 \p_2 - M_2 \p_1 ) M_3 - (\p_2 M_1 - \p_1 M_2) M_3 \right]\,dx dy,
\end{split}
\end{equation}
where $M_s$ is the saturation magnetization, $A$ is the exchange constant, $K$ the anisotropy constant,
and $D$ the DM constant.
Spatial derivatives in Eq.~\eqref{eq:energy0} are denoted by $\p_\mu$ with $\mu=1,2$ and
$\p_1=\p_x, \p_2=\p_y$.
We have not included the energy of the demagnetizing field in Eq.~\eqref{eq:energy0} 
because it does not affect skyrmion configurations in a qualitatively significant way \cite{BegChernyshenko_arXiv2014};
it introduces a dependence of the skyrmion size on the film thickness \cite{KiselevBogdanov_JPD2011}.
Note that $\Energy$ in Eq.~\eqref{eq:energy0} is actually the energy per unit length along the easy axis
perpendicular to the film.

We define the normalised magnetization
$\bm{m} = \bm{M}/M_s$, so that $\bm{m}^2=1$.
We further use $\ldm = 2A/|D|$ as the unit of length,
hence the energy measured in units of $2A$ is given by (see also \cite{WilsonButenko_PRB2014})
\begin{equation}  \label{eq:energy}
\begin{split}
\Energy & = \frac{1}{2} \int \p_\mu \bm{m}\cdot \p_\mu \bm{m}\,dx dy + \frac{\anisotropy}{2} \int (m_1^2+m_2^2)\,dx dy  \\
                &  + \dm \int \left[ (m_1 \p_2 - m_2 \p_1) m_3 - (\p_2 m_1 - \p_1 m_2) m_3 \right]\,dx dy,
\end{split}
\end{equation}
where
\begin{equation}  \label{eq:dm_parameter}
\dm = \frac{D}{|D|} = \pm 1
\end{equation}
will be referred to as the chirality and
\begin{equation}   \label{eq:anisotropy}
\anisotropy \equiv \frac{\Anisotropy}{\Anisotropy_0},\qquad \Anisotropy_0 = \frac{D^2}{4A}
\end{equation}
is the rationalized (dimensionless) anisotropy constant.
Unless otherwise stated, we choose chirality $\dm=1$ in all of our numerical calculations,
while $\anisotropy$ is taken to be positive (easy-axis anisotropy).

The conservative Landau-Lifshitz (LL) equation associated with the energy \eqref{eq:energy} is
\begin{align}  \label{eq:LL}
\frac{\p \bm{m}}{\p t} & = -\bm{m} \times \bm{f},  \\
\bm{f} \equiv -\frac{\delta \Energy}{\delta \bm{m}} & = \Delta\bm{m} + \anisotropy m_3 \ez  \notag \\
  & - 2\dm\,\left[ \p_2 m_3\, \ex - \p_1 m_3\,\ey + (\p_1 m_2 - \p_2 m_1)\,\ez \right].  \notag
\end{align}
The time variable $t$ is measured in units of $\tau_0=2A M_s/(\gamma \DM^2)$ where
$\gamma$ is the gyromagnetic ratio.

We recall at this point that magnetic configurations are characterised by
the skyrmion number defined as
\begin{equation}  \label{eq:skyrmion_number}
\Skyrmion = \frac{1}{4\pi} \int \skyrmion\, dx dy,\qquad 
\skyrmion = \frac{1}{2}  \emn \bm{m}\cdot(\p_\nu\bm{m}\times \p_\mu\bm{m}),
\end{equation}
where $\skyrmion$ is called the {\it topological density}.
The skyrmion number $\Skyrmion$ is integer-valued ($\Skyrmion=0,\pm 1,\pm 2,\ldots$)
for all magnetic configurations such that $\bm{m}=(0,0,\pm 1)$ at spatial infinity.
For definiteness we assume $\bm{m}=(0,0,1)$ in all our calculations.

We also construct a tensor $\sigma_{\mu\nu}$ defined from
\begin{equation}  \label{eq:sigma_divergence}
\p_\nu \sigma_{\mu\nu} = -\bm{f}\cdot \p_\mu \bm{m} = \frac{\delta \Energy}{\delta \bm{m}}\cdot\p_\mu\bm{m}.
\end{equation}
A formal calculation gives the tensor components
\begin{equation}  \label{eq:sigma}
\begin{split}
\sigma_{11} = & \frac{1}{2} \left( \p_1\bm{m}\cdot \p_1\bm{m} -  \p_2\bm{m}\cdot \p_2\bm{m} \right) 
  + \frac{\anisotropy}{2} (m_1^2 + m_2^2)   \\
   & + \lambda (m_1 \p_2 m_3 - m_3 \p_2 m_1 )  \\
\sigma_{12} = & -\p_1\bm{m}\cdot \p_2\bm{m} + \lambda (m_3 \p_1 m_1 - m_1 \p_1 m_3 )  \\
\sigma_{21} = & -\p_1\bm{m}\cdot \p_2\bm{m} + \lambda (m_2 \p_2 m_3 - m_3 \p_2 m_2 )  \\
\sigma_{22} = & \frac{1}{2} \left( \p_2\bm{m}\cdot \p_2\bm{m} - \p_1\bm{m}\cdot \p_1\bm{m} \right) 
 +  \frac{\anisotropy}{2} (m_1^2 + m_2^2)   \\
  & + \lambda (m_3 \p_1 m_2 - m_2 \p_1 m_3 ).
\end{split}
\end{equation}
The topological density $\skyrmion$ together with the tensor $\sigma_{\mu\nu}$ provide
important theoretical tools for the analysis of both static and dynamical properties
of the Landau-Lifshitz equation \cite{PapanicolaouTomaras_NPB1991,KomineasPapanicolaou_PhysD1996}.

\section{Static skyrmions}
\label{sec:skyrmions}

The uniform (ferromagnetic) states $\bm{m}=(0,0\pm 1)$ are the simplest static solutions 
of the LL equation \eqref{eq:LL} with total energy $\Energy =0$.
For large rationalised anisotropy $\anisotropy$ the uniform state is the ground state of the system.
However, for sufficiently low anisotropy a {\it spiral state} with energy $\Energy <  0$ becomes the ground state \cite{BogdanovHubert_JMMM1994}.
The transition happens at the critical value
\begin{equation}
\anisotropy_c = \frac{\pi^2}{4} \approx 2.4674.
\end{equation}
The period of the spiral increases for increasing anisotropy and goes to infinity at the critical value $\anisotropy_c$.

In order to find nontrivial static solutions which satisfy the equation $\bm{m}\times\bm{f}=0$
we may employ a relaxation algorithm based on
\begin{equation}  \label{eq:relaxation}
\frac{\p \bm{m}}{\p t} = - \bm{m} \times (\bm{m} \times \bm{f}),
\end{equation}
typically implemented on a $300\times 300$ lattice with uniform spacing $\Delta x = \Delta y = 0.1$
and Neumann boundary conditions.
The numerical lattice is large enough so that it can be assumed that an infinite film is simulated
 in all presented calculations.
An initial spin configuration will evolve under Eq.~\eqref{eq:relaxation} in such a way that its energy
decreases monotonically and eventually converges to a static solution of Eq.~\eqref{eq:LL}. 

We now focus on axially symmetric skyrmion configurations.
These are conveniently described in terms of the standard spherical parametrisation for the magnetization given by
\begin{equation}
m_1 = \sin\Theta \cos\Phi,\quad m_2 = \sin\Theta \sin\Phi,\quad m_3 = \cos\Theta
\end{equation}
using the ansatz
\begin{equation}  \label{eq:axially_symmetric}
\Theta = \theta(\rho),\qquad \Phi = \phi + \pi/2,
\end{equation}
where $(\rho,\phi)$ are polar coordinates.
Stationary solutions of the energy functional \eqref{eq:energy0} then satisfy the ordinary
differential equation \cite{BogdanovHubert_JMMM1999,WilsonButenko_PRB2014}
\begin{equation}  \label{eq:LL_staticAxial}
\frac{d^2\theta}{d\rho^2} + \frac{1}{\rho}\frac{d\theta}{d\rho} - \left(\anisotropy
+\frac{1}{\rho^2} \right) \cos\theta \sin\theta + \frac{2\dm}{\rho}\,\sin^2\theta = 0
\end{equation}
while the skyrmion number defined from Eq.~\eqref{eq:skyrmion_number} reduces to
\begin{equation}  \label{eq:Skyrmion_axial}
\Skyrmion = \frac{1}{2} \int_0^\infty \frac{d m_3}{d \rho}\, d\rho = \frac{1}{2} \left[m_3(\infty) - m_3(0) \right]
\end{equation} 
where $m_3=\cos\theta=m_3(\rho)$ is the third component of magnetization.
Thus, if Eq.~\eqref{eq:Skyrmion_axial} is solved with boundary conditions
$\theta(\rho=0)=\pi$ and $\theta(\rho\to\infty)=0$,
it leads to a static skyrmion with $\Skyrmion=1$.

\begin{figure}[t]
\begin{center}
\includegraphics[width=6truecm]{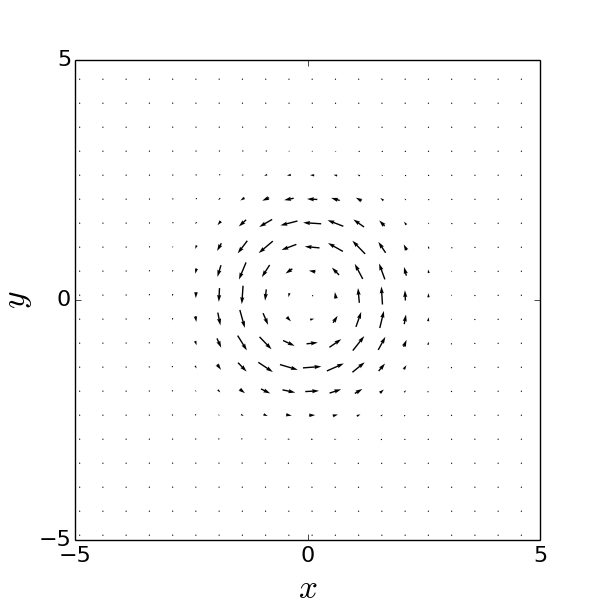}
\caption{The static axially symmetric ($\Skyrmion=1$) skyrmion represented through the projection
$(m_1,m_2)$ of the magnetization vector on the plane
for anisotropy $\anisotropy=3$.}
\label{fig:skyrmion_vecs}
\end{center}
\end{figure}

Boundary-value problems of the above nature are typically solved by some sort of a shooting method
\cite{BogdanovHubert_JMMM1999}.
Here we employ a more general (actually simpler) method based on the fully-dissipative algorithm \eqref{eq:relaxation}
which does not a priori assume axial symmetry.
Instead, we may initialize Eq.~\eqref{eq:relaxation} with an essentially arbitrary spin configuration with $\Skyrmion=1$
which then converges to a static $\Skyrmion=1$ skyrmion that is a local minimum
of the energy functional \eqref{eq:energy}.
The result is shown in Fig.~\ref{fig:skyrmion_vecs} for anisotropy $\anisotropy=3$ and its axial symmetry is evident.
Also note that the chirality of the skyrmion (counter-clockwise rotation in Fig.~\ref{fig:skyrmion_vecs})
is fixed by the sign of the DM constant ($\dm=1$).
If we choose $\dm=-1$ then the profile $m_3=m_3(\rho)$ of the skyrmion remains unchanged
but its chirality is reversed.
To be sure, the skyrmion number is $\Skyrmion=1$ for either choice of chirality.

\begin{figure}[t]
\begin{center}
\includegraphics[width=6truecm]{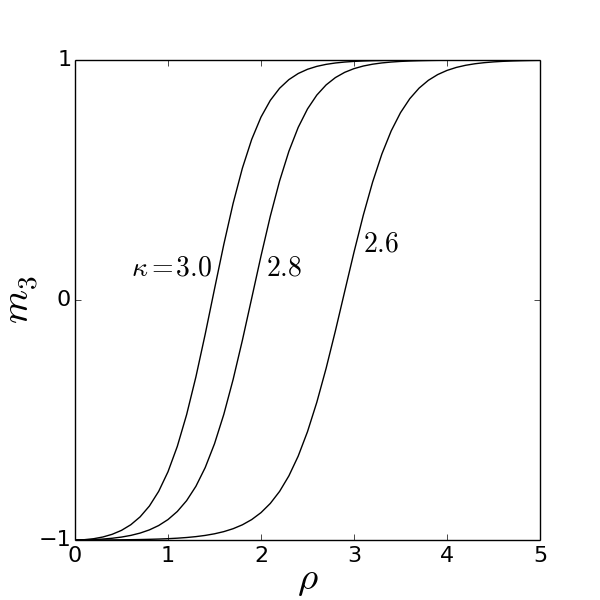}
\caption{The profiles $m_3=\cos\theta=m_3(\rho)$ of skyrmions ($\Skyrmion=1$) for three values
 of the anisotropy parameter $\anisotropy=2.6, 2.8, 3.0$.}
\label{fig:skyrmion_profiles}
\end{center}
\end{figure}

Fig.~\ref{fig:skyrmion_profiles} shows the skyrmion profiles for three values of the anisotropy constant $\anisotropy$.
As $\anisotropy$ approaches $\anisotropy_c$ from above, the radius of the skyrmion diverges to infinity.
There is no evidence for the existence of skyrmions when $\anisotropy < \anisotropy_c$.

Other axially symmetric static skyrmions can be found if we assume multiple rotations of the magnetization
as we move radially from the skyrmion center.
These were first identified in Ref.~\cite{BogdanovHubert_JMMM1999} solving Eq.~\eqref{eq:LL_staticAxial} 
with boundary conditions $\theta(\rho=0)=k\pi, \theta(\rho\to\infty)=0$, with $k=1,2,3,\ldots$,
where $k=1$ is the fundamental ($\Skyrmion=1$) skyrmion already discussed,
$k=2$ leads to a $\Skyrmion=0$ configuration, in view of Eq.~\eqref{eq:Skyrmion_axial}, and so on.

\begin{figure}[t]
\begin{center}
\includegraphics[width=6truecm]{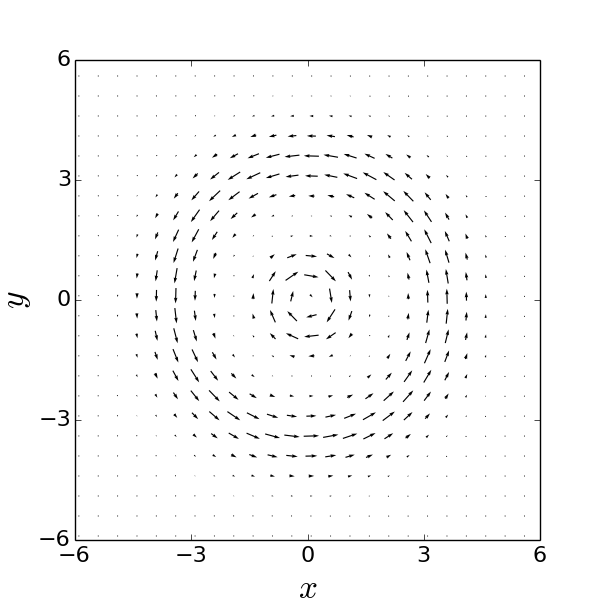}
\caption{The static axially symmetric skyrmionium ($\Skyrmion=0$) represented through the projection
$(m_1,m_2)$ of the magnetization vector on the plane
for anisotropy $\anisotropy=3$.}
\label{fig:skyrmionium_vecs}
\end{center}
\end{figure}

\begin{figure}[t]
\begin{center}
\includegraphics[width=6truecm]{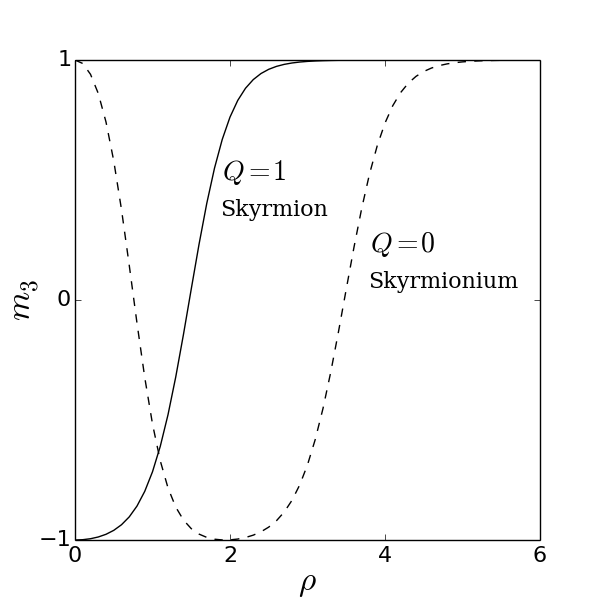}
\caption{The profiles $m_3=\cos\theta=m_3(\rho)$ of a skyrmion ($\Skyrmion=1$) and a skyrmionium ($\Skyrmion=0$) 
for anisotropy $\anisotropy=3$.}
\label{fig:skyrmum_profiles}
\end{center}
\end{figure}

In order to apply our relaxation algorithm for $k=2$, we first construct a suitable ansatz to be used
as initial condition in Eq.~\eqref{eq:relaxation}.
We recall the axially symmetric $\Skyrmion=1$ configuration constructed above, 
which we now denote by $\bm{n}=(n_1,n_2,n_3)$.
We then apply the transformation \cite{KomineasPapanicolaou_NL1998}
\begin{equation}  \label{eq:ansatz_Q0}
m_1 = 2 n_3 n_1,\qquad m_2 = 2 n_3 n_2,\qquad m_3 = 2 n_3^2 - 1.
\end{equation}
The resulting configuration $\bm{m}$ remains axially symmetric
but its skyrmion number is $\Skyrmion=0$.
This follows from the fact that $m_3(\rho=0)=1=m_3(\rho\to\infty)$.
The magnetization rotates to $m_3=-1$ at some intermediate radius.
We apply algorithm \eqref{eq:relaxation} inserting configuration \eqref{eq:ansatz_Q0} as initial condition.
The algorithm converges to an axially symmetric configuration with $\Skyrmion=0$, 
shown in Fig.~\ref{fig:skyrmionium_vecs}, for $\anisotropy=3$.
Such a configuration may be called a {\it ``skyrmionium''} \cite{FinazziSavoini_PRL2013}
because it consists internally of a skyrmion and an antiskyrmion.
A comparison between the skyrmion and the skyrmionium is given through their profiles $m_3(\rho)$
plotted in Fig.~\ref{fig:skyrmum_profiles} for the same anisotropy parameter $\anisotropy=3$.

The existence of a static chiral skyrmion has been rigorously established within a related model
only for $\Skyrmion=1$ \cite{Melcher_PRSB2014}.
We have searched numerically for static solutions of Eq.~\eqref{eq:LL} with $\Skyrmion=-1$ or $2$,
which lack axial symmetry, but our relaxation algorithm \eqref{eq:relaxation} did not converge to any such
static solutions.
We will continue this study focusing on the axially symmetric configurations shown 
in Figs.~\ref{fig:skyrmion_vecs}, \ref{fig:skyrmionium_vecs}.

For static solutions such as those discussed in the present section 
the tensor of Eq.~\eqref{eq:sigma} satisfies $\p_\lambda \sigma_{\mu\lambda} = 0$.
This yields $x_\nu \p_\lambda \sigma_{\mu\lambda} = 0$, while integration of both sides over the
entire plane and an elementary application of the divergence theorem leads to \cite{PapanicolaouSpathis_NL1999}
\begin{equation}  \label{eq:virial_static}
\int \sigma_{\mu\nu}\, dx dy=0
\end{equation}
where the indices $\mu$ and $\nu$ take the values $1$ or $2$ in any combination and thus lead 
to four independent virial relations that must be satisfied by any static solution.
A special combination of these relations yields the result
\begin{equation}
\int (\sigma_{11} + \sigma_{22})\, dx dy = 0 \Rightarrow 2\Ea + \Edm = 0,
\end{equation}
which was obtained through a scaling argument in Ref.~\cite{BogdanovHubert_PSS1994} in analogy to
Derrick-Hobart relations \cite{Derrick_JMP1964,Hobart_PPS1963}.
Here, $\Ea$ is the positive anisotropy energy while $\Edm$ is the Dzyaloshinskii-Moriya energy
that may be positive or negative and thus does not exclude nontrivial static solutions.
Another special case of Eqs.~\eqref{eq:virial_static} gives
\begin{align}
& \int (\sigma_{12} - \sigma_{21}) dx dy = 0 \Rightarrow \notag \\
& \int \left[ m_3 (\bnabla\cdot \bm{m}) - (\bm{m}\cdot\bnabla) m_3 \right] dx dy = 0,
\end{align}
where $\bnabla = \ex \p_1 + \ey \p_2$, and it is entirely due to the anisotropic DM interaction.

\section{Skyrmion dynamics}
\label{sec:dynamics}

We now study the dynamics associated with the two static solutions calculated in Sec.~\ref{sec:skyrmions};
namely, the topological ($\Skyrmion=1$) skyrmion and the non-topological ($\Skyrmion=0$) skyrmionium.
We shall find that significant differences arise in the two cases,
as anticipated by an important link between topology and dynamics established in our early work
\cite{PapanicolaouTomaras_NPB1991,KomineasPapanicolaou_PhysD1996} through the construction
of unambiguous conservation laws.

The main result is established rigorously by examining the time evolution of the topological density $\skyrmion$
of Eq.~\eqref{eq:skyrmion_number}.
A straightforward application of the LL equation \eqref{eq:LL} leads to
\begin{equation}  \label{eq:vorticity_derivative}
\dot{\skyrmion} = -\emn \p_\mu (\bm{f}\cdot\p_\nu\bm{m}) = \emn\,\p_\mu\p_\lambda \sigma_{\nu\lambda}
\end{equation}
where the overdot denotes time derivative and $\sigma_{\nu\lambda}$ is the tensor defined
in Eq.~\eqref{eq:sigma}.
An immediate consequence of Eq.~\eqref{eq:vorticity_derivative} is that the integrated topological density,
the skyrmion number $\Skyrmion$ of Eq.~\eqref{eq:skyrmion_number} is conserved, as expected.
Furthermore, the appearance of the double spatial derivative in the right-hand side of
Eq.~\eqref{eq:vorticity_derivative} suggests that some of the low moments of the topological density
are also conserved.
The lowest nontrivial moments are given by
\begin{equation}  \label{eq:vorticity_moments}
I_\mu = \int x_\mu\skyrmion\, dx dy,\qquad \mu=1,2
\end{equation}
and their sconservation ($\dot{I}_\mu=0$) is demonstrated by a simple application of Eq.~\eqref{eq:vorticity_derivative}
and the divergence theorem.
In order to reveal the physical content of moments \eqref{eq:vorticity_moments} we note that
under a rigid translation of spatial coordinates by a constant vector ($x_\mu \to x_\mu + c_\mu$)
the moments transform as
\begin{equation}  \label{eq:translation}
I_\mu \to I_\mu + 4\pi\Skyrmion\, c_\mu.
\end{equation}
which indicate an important difference between topological ($\Skyrmion \neq 0$) and
non-topological ($\Skyrmion = 0$) magnetic solitons.
The two cases are studied in turn in subsections \ref{sec:hall_motion} and \ref{sec:newtonian_motion}.

In order to probe skyrmion dynamics we consider the effect of an applied magnetic field
$\bhext = \bhext(x,y,t)$ which may be a nontrivial function of both spatial and time variables.
The question is then to predict the behavior of the magnetic configuration after the field is turned on.
The Landau-Lifshitz equation \eqref{eq:LL} is then modified by the simple replacement $\bm{f} \to \bm{f} + \bhext$.
In particular, relation \eqref{eq:vorticity_derivative} becomes
\begin{equation}
\dot{\skyrmion} = \emn\,\p_\mu\p_\lambda \sigma_{\nu\lambda} - \emn \p_\mu (\bhext\cdot\p_\nu \bm{m}).
\end{equation}
As a result the skyrmion number $\Skyrmion$ is still conserved, as expected, 
but the moments $I_\mu$ are no longer conserved and satisfy
\begin{equation}  \label{eq:momentum_dynamics}
\dot{I}_\mu = \emn \int \bhext\cdot \p_\nu \bm{m}\, dx dy.
\end{equation}
Yet studying the degree to which the conservation is violated will give important information
on the motion of skyrmions in the presence of the applied field.

The rationalized field $\bhext$ is related to the physical field $\bm{H}$ by
\begin{equation}  \label{eq:hext}
\bhext = \frac{\bm{H}}{H_0}\qquad H_0=\frac{D^2}{2A \mu_0 M_s}
\end{equation}
and we may also write $H_0=M_s(\lex/\ldm)^2$
where $\lex=\sqrt{2A/(\mu_0 M_s^2)}$ is the exchange length, the most commonly used length scale in micromagnetics.

\subsection{Hall motion of $\Skyrmion=1$ skyrmion}
\label{sec:hall_motion}

In view of Eq.~\eqref{eq:translation} the physical interpretation of moments $I_\mu$ in Eq.~\eqref{eq:vorticity_moments} 
depends crucially on the skyrmion number $\Skyrmion$.
For $\Skyrmion \neq 0$, the normalized moments
\begin{equation}  \label{eq:guiding_center}
R_\mu = \frac{I_\mu}{4\pi\Skyrmion} = \frac{1}{4\pi\Skyrmion} \int x_\mu\skyrmion\, dx dy,
\end{equation}
provide a measure of position.
The 2D vector $\bm{R}=(R_x,R_y)$ will be referred to as the {\it guiding center} of the magnetic configuration
in question and is conserved in the absence of a magnetic field gradient.
Thus, a magnetic soliton with $\Skyrmion \neq 0$ cannot move freely and is spontaneously pinned
within the ferromagnetic medium.
However, motion is possible in the presence of an applied magnetic field gradient.

We consider an initially static skyrmion such as the $\Skyrmion=1$ skyrmion shown in Fig.~\ref{fig:skyrmion_vecs},
and apply an applied magnetic field with a gradient in the $x$ direction:
\begin{equation}  \label{eq:field_gradient}
\bhext = (0,0,\hext),\qquad \hext = g\,x e^{-x^2/a^2}
\end{equation}
where $g$ is the strength of the gradient and $a$ is a constant.
The field gradient is almost uniform for $x < a$ and it fades out for $x > a$.
A pure gradient ($\hext = gx$) may be achieved in the formal limit $a\to\infty$.
However, such a limit should be taken with caution because the field would then reach high
values at large $x$ and might destroy the uniform ground state $\bm{m}=(0,0,1)$
in the presence of dissipation.

The velocity of the skyrmion guiding center is obtained by inserting Eq.~\eqref{eq:field_gradient}
in Eq.~\eqref{eq:momentum_dynamics} and applying a partial integration
\begin{equation}  \label{eq:skyrmion_velocity0}
\dot{R}_x = 0,\quad \dot{R}_y = -\frac{1}{4\pi\Skyrmion} \int \p_x \hext\, (1-m_3)\, dx dy.
\end{equation}
The guiding center drifts along the $y$ axis, thus in a direction perpendicular to the applied field gradient.
For a pure gradient $\hext=gx$, applying Eq.~\eqref{eq:skyrmion_velocity0} we find the simple formula
\begin{equation}  \label{eq:skyrmion_velocity}
\dot{R}_x = 0,\qquad \dot{R}_y= -\frac{g\tmagn}{4\pi\Skyrmion},
\end{equation}
where $\tmagn=\int (1-m_3)\, dx dy$ is the total magnetization.
For the static $\Skyrmion=1$ skyrmion of Fig.~\ref{fig:skyrmion_vecs} the numerical calculation
gives $\mu=15.1$,  thus the expected velocity is $\dot{R}_y = -1.20g$.
The above prediction is rigorously correct during the initial stages of the process,
but deviations from a constant velocity are possible at later stages
because the moment $\tmagn$ is not by itself conserved.

\begin{figure}[t]
\begin{center}
\includegraphics[width=4.2truecm]{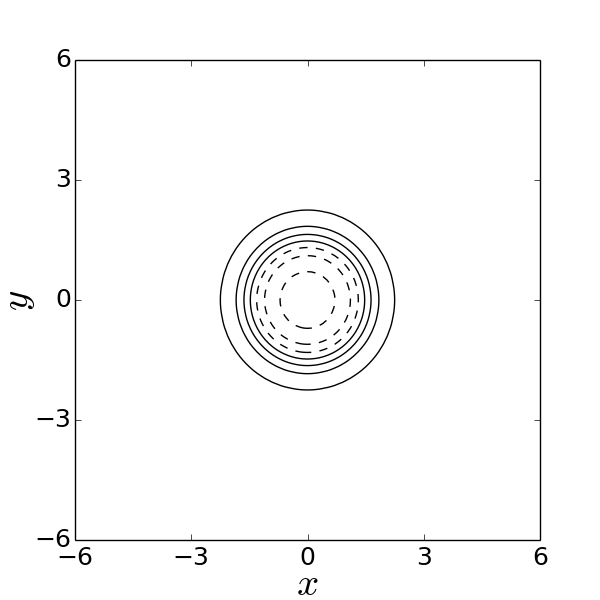}  
\includegraphics[width=4.2truecm]{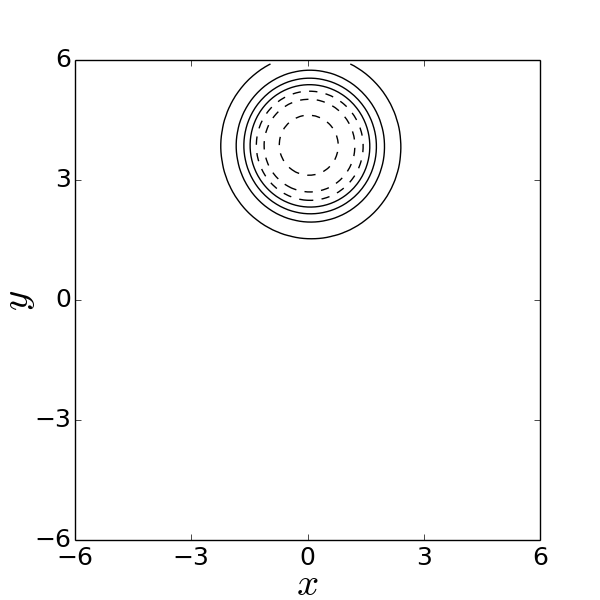}  
\caption{Contour plots for $m_3$ for a skyrmion under the field-gradient \eqref{eq:field_gradient} with $g=-0.1,\, a=10$.
(Left) The static $\Skyrmion=1$ skyrmion of Fig.~\ref{fig:skyrmion_vecs} is placed at the origin at initial time $t=0$.
(Right) The skyrmion at simulation time $t=30$.
It moves along the $y$ axis perpendicular to the field gradient.
The contour levels plotted are $m_3=0.9,0.6,0.3,0.0$ (solid lines) and $m_3=-0.3,-0.6,-0.9$ (dashed lines).
} 
\label{fig:skyrmion_vecs_g0100t30}
\end{center}
\end{figure}

\begin{figure}[t]
\begin{center}
\includegraphics[height=6.0truecm]{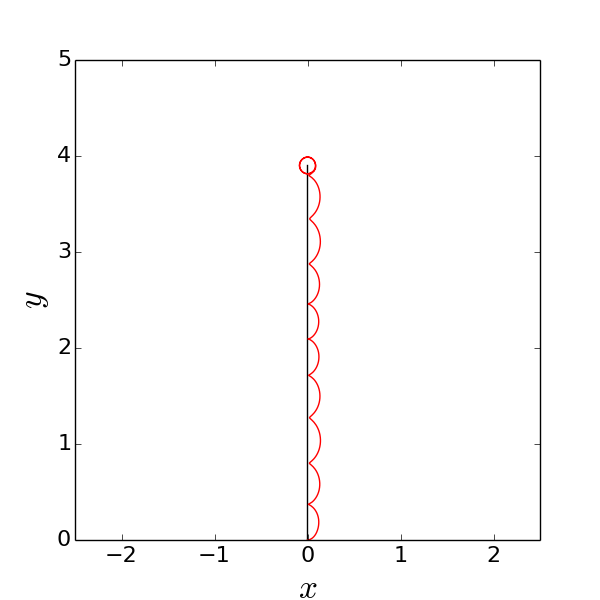}
\includegraphics[height=6.0truecm]{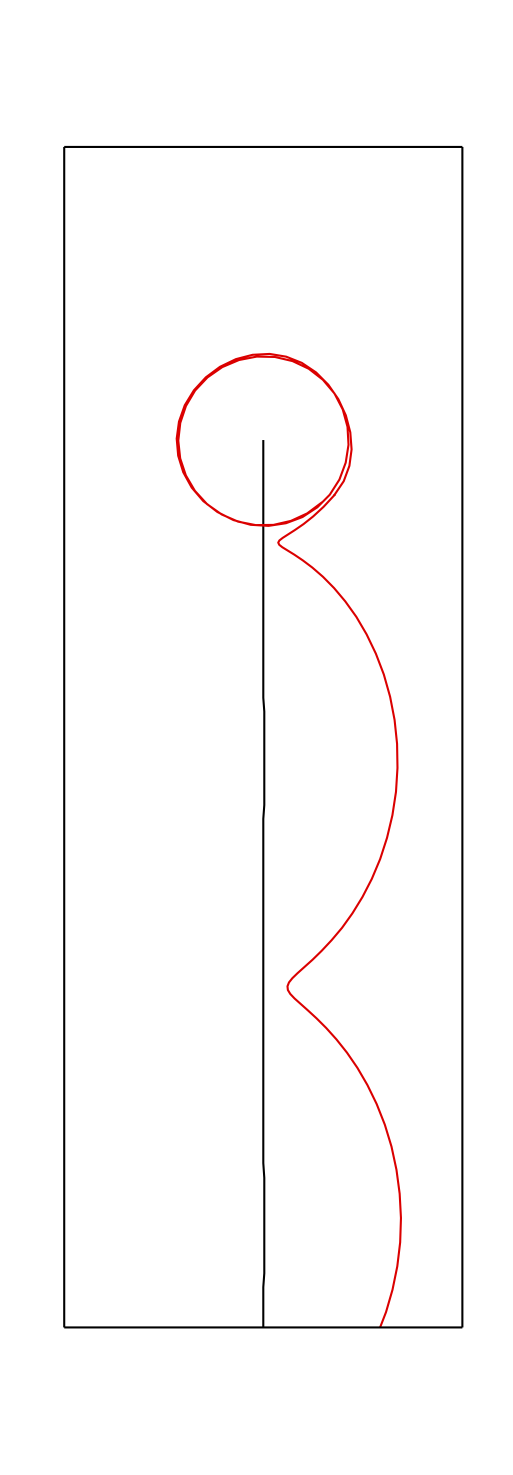}
\caption{(Left) The trajectory of a $\Skyrmion=1$ skyrmion under the influence 
of the field-gradient \eqref{eq:field_gradient} with  $g=-0.1,\; a=10$.
The guiding center propagates along the $y$-axis; thus in a direction perpendicular
to the applied field-gradient (straight black line).
The trajectory for the $(\xmagn, \ymagn)$ of Eq.~\eqref{eq:moments_magnetization} is
the cycloid shown by a solid red line.
After the applied field is suddenly switched off the guiding center ceases to move further
and the skyrmion trajectory organizes itself in a rotational cyclotron-type motion
around the fixed guiding center.
(Right) A magnification of the latest stages of the process is shown.} 
\label{fig:Q1_g0100_trajectory}
\end{center}
\end{figure}

The preceding results are verified by a direct numerical simulation.
We apply a field gradient \eqref{eq:field_gradient} with $g=-0.1$ and $a=10$
and solve the initial value problem with the initial condition provided by a static $\Skyrmion=1$ skyrmion 
(with $\anisotropy=3$) placed at the origin $(0,0)$.
The skyrmion moves coherently in the $y$ direction, that is, 
perpendicular to the field gradient.
Fig.~\ref{fig:skyrmion_vecs_g0100t30} shows the skyrmion at time $t=0$ and at $t=30$
when it has drifted approximately $\Delta y = 4$ units along the $y$ axis.
The guiding center, shown in Fig.~\ref{fig:Q1_g0100_trajectory} (solid black line),
moves along the $y$ axis with an almost constant velocity. 
Its initial ($t<4$) velocity is $\dot{R}_y=0.12$ in excellent agreement with the
prediction obtained from Eq.~\eqref{eq:skyrmion_velocity}.
At later times the guiding center velocity increases to $\dot{R}_y=0.13$ while a corresponding
increase in the moment $\tmagn$ is observed.

One could heuristically invent more quantities to describe the skyrmion position,
whose usefulness depends on their connection to actual measurements.
A plausible measure of position is given by the moments of the third component of the magnetization:
\begin{equation}  \label{eq:moments_magnetization}
\xmagn = \frac{\int x(1-m_3)\, dx dy}{\int (1-m_3)\, dx dy},\quad
\ymagn = \frac{\int y(1-m_3)\, dx dy}{\int (1-m_3)\, dx dy}.
\end{equation}
Fig.~\ref{fig:Q1_g0100_trajectory} also shows the trajectory $(\xmagn, \ymagn)$ 
obtained through the numerical simulation.
The position $(\xmagn, \ymagn)$ closely follows the guiding center, but is also decorated by Larmor oscillations.
The picture is strongly reminiscent of the cycloidal Hall motion of an electric charge that
moves under the influence of an electric field and a magnetic field
perpendicular to the plane of the motion (see Ref.~\cite{LandauLifshitz_Fields} pp 57). 

The preceding analogy with Hall motion is further substantiated by suddenly switching off the applied
magnetic field gradient after the skyrmion has moved a distance, say, $\Delta y = 4$ along the $y$-axis.
The guiding center then ceases to move further, while the skyrmion organizes itself in a cyclotron-type
rotational motion around the pinned guiding center
(see Fig.~\ref{fig:Q1_g0100_trajectory}).

\subsection{Newtonian motion of $\Skyrmion=0$ skyrmionium}
\label{sec:newtonian_motion}

According to Eq.~\eqref{eq:translation} the moments $I_\mu$ are invariant under rigid translations
for a $\Skyrmion=0$ skyrmion.
Therefore, unlike the case of the previous subsection for $\Skyrmion \neq 0$, 
the conservation laws for $I_\mu$ do not exclude the possibility of free translational motion
when $\Skyrmion = 0$.
Furthermore, the {\it linear momentum} is defined via the conserved $I_\mu$ as 
\cite{PapanicolaouTomaras_NPB1991,KomineasPapanicolaou_PhysD1996}
\begin{equation} \label{eq:linear_momentum}
P_\mu = \emn I_\nu,\qquad \mu, \nu = 1\; \hbox{or}\; 2.
\end{equation}
The invariance of linear momentum under rigid translations is in accordance 
to the properties of ordinary momentum of a point particle in Newtonian dynamics.

\begin{figure}[t]
\begin{center}
\includegraphics[width=4.2truecm]{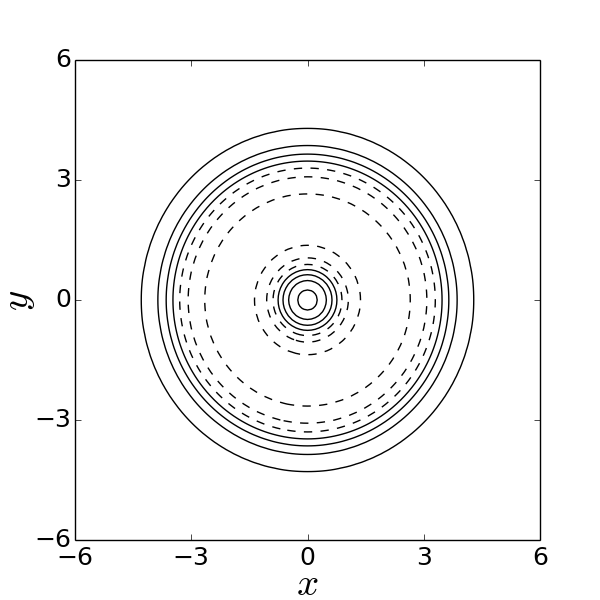}\hspace{0pt}
\includegraphics[width=4.2truecm]{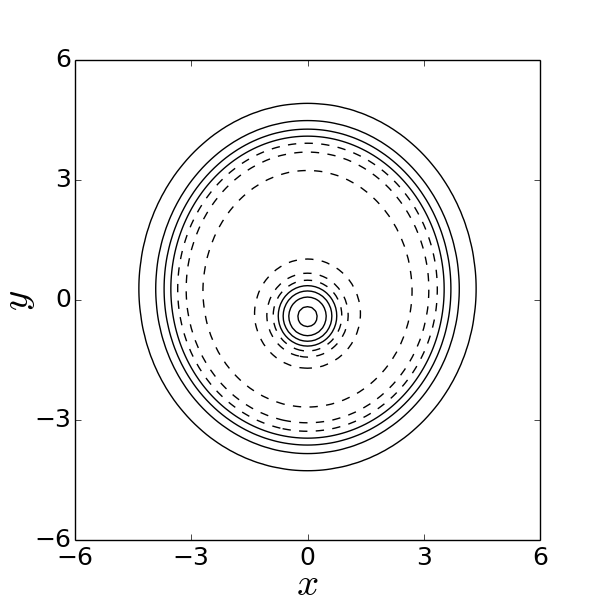}
\caption{Contour plot for $m_3$ for the static skyrmionium (left) and the propagating skyrmionium
with velocity $v=0.07$ (right). The contour levels plotted with solid lines are $m_3=0.9,0.6,0.3,0.0$ and
with dashed lines $m_3=-0.3,-0.6,-0.9$.}
\label{fig:skyrmionium_vel_m3}
\end{center}
\end{figure}

Newtonian dynamics allows for a steady-state with constant velocity when no forces are applied.
We thus look for steady-states propagating rigidly with constant velocity $\bm{v}=(v_1, v_2)$
i.e., $\bm{m} = \bm{m}(\bm{x}-\bm{v} t; \bm{v})$ satisfying
$\dot{\bm{m}} = -v_\lambda \p_\lambda\bm{m}$ and thus
\begin{equation}  \label{eq:LL_steadystate}
v_\lambda \p_\lambda\bm{m} = \bm{m}\times \bm{f}.
\end{equation}
Now, take the cross product of both sides with $\bm{m}$, then the dot product with $\p_\nu\bm{m}$,
to obtain
\begin{equation}  \label{eq:virial_differential}
\epsilon_{\nu\lambda} v_\lambda \skyrmion = \p_\lambda \sigma_{\nu\lambda}
\end{equation}
where $\skyrmion$ is the topological density defined in Eq.~\eqref{eq:skyrmion_number}
and $\sigma_{\nu\lambda}$ the tensor defined from Eq.~\eqref{eq:sigma}.
Eq.~\eqref{eq:virial_differential} provides the basis for the derivation of  a series
of interesting virial relations generalizing the virial theorems \eqref{eq:virial_static}
satisfied by static solutions.

An immediate consequence of Eq.~\eqref{eq:virial_differential} is obtained by integrating
both sides over the entire $xy$-plane:
\begin{equation}  \label{eq:virial0}
\epsilon_{\nu\lambda} v_\lambda \Skyrmion=0
\end{equation}
where $\Skyrmion$ is the skyrmion number of the magnetic soliton in question.
Therefore, we conclude that topological solitons ($\Skyrmion \neq 0$) cannot be found
in rigid translational motion ($v_\lambda = 0$).
This elementary result is in agreement with a similar conclusion reached for
($\Skyrmion = 1$) skyrmions on the basis of the conservation of the guiding center $\bm{R}$
in the absence of an applied field gradient.

On the other hand, Eq.~\eqref{eq:virial0} does not exclude a rigidly moving non-topological soliton
($\Skyrmion=0$) because it is then trivially satisfied for any velocity $v_\lambda$.
In this case, Eq.~\eqref{eq:virial_differential} can be further iterated by multiplying
both sides with $x_\mu$ and then integrating over the entire $xy$-plane to obtain
the virial relations \cite{PapanicolaouSpathis_NL1999}
\begin{equation}
(v_\lambda P_\lambda) \delta_{\mu\nu} - P_\mu v_\nu = \int \sigma_{\mu\nu}\, dx dy
\end{equation}
where $P_\mu$ is the linear momentum defined in Eq.~\eqref{eq:linear_momentum}
while indices $\mu$ and $\nu$ take the values $1$ or $2$ in any combination
and thus lead to four independent virial relations.
These must be satisfied by any non-topological ($\Skyrmion=0$) magnetic soliton
moving rigidly with constant velocity $\bm{v}=(v_1,v_2)$.

Actual solutions of Eq.~\eqref{eq:LL_steadystate} are obtained 
numerically by a generalization of the relaxation algorithm \eqref{eq:relaxation}
used for the calculation of static solutions in Sec.~\ref{sec:skyrmions}.
For definiteness we assume rigid motion along the $x$-axis $(v_1=v, v_2=0)$.
Eq.~\eqref{eq:relaxation} is then generalized according to
\begin{equation}  \label{eq:relaxation_traveling}
\frac{\p \bm{m}}{\p t} = -\bm{m} \times \left(\bm{m}\times \bm{f}-v\, \frac{\p \bm{m}}{\p x} \right),\quad
v = u - P
\end{equation}
where the velocity $v$ is determined self-consistently in terms of an arbitrary input parameter $u$
and the momentum integral $P = P_x = I_2$ calculated from Eq.~\eqref{eq:linear_momentum} for
each time step.
It should be noted that the momentum is not conserved by the dissipative dynamics
\eqref{eq:relaxation_traveling}, hence the ``velocity'' $v$ evolves together with the spin configuration
until they both reach definite terminal values that are a local minimum of the
Lyapunov functional $F=\Energy +  \frac{1}{2}(u-P)^2$.
In view of the constraint $\bm{m}^2=1$, which is compatible with Eq.~\eqref{eq:relaxation_traveling},
the terminal state will satisfy the differential equation
\begin{equation}
v\,\frac{\p \bm{m}}{\p x} = \bm{m}\times\bm{f}
\end{equation}
and thus describes a magnetic soliton that moves rigidly along the $x$-axis with constant velocity $v$
which depends on the input parameter $u$ and may be varied accordingly.

The algorithm is initiated with the static skyrmionium of Fig.~\ref{fig:skyrmionium_vecs}
and a nonzero value for the input parameter $u$.
The algorithm converges to a steady-state configuration with velocity in the range
\begin{equation}  \label{eq:skyrmionium_velocity}
0 \leq v < v_c,\qquad v_c \approx 0.102.
\end{equation}
Fig.~\ref{fig:skyrmionium_vel_m3} shows the static together with a propagating skyrmionium via contour plots for
the magnetization component $m_3$.
Note that a skyrmionium with $v \neq 0$ is no longer axially symmetric.

\begin{figure}[t]
\begin{center}
\includegraphics[width=4.2truecm]{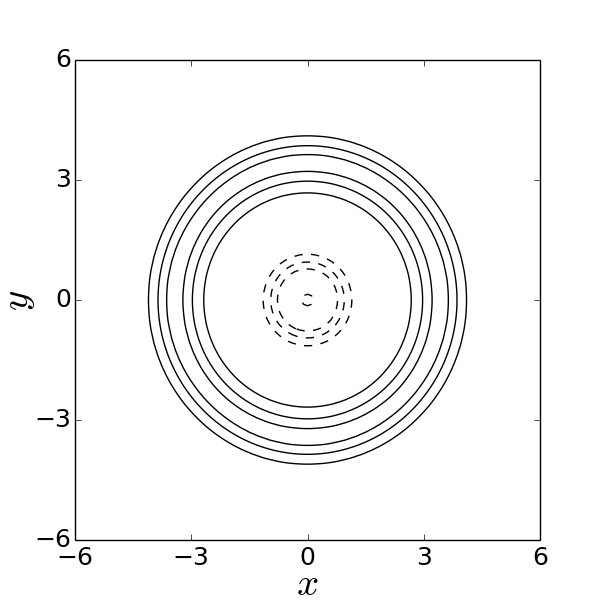}\hspace{0pt}
\includegraphics[width=4.2truecm]{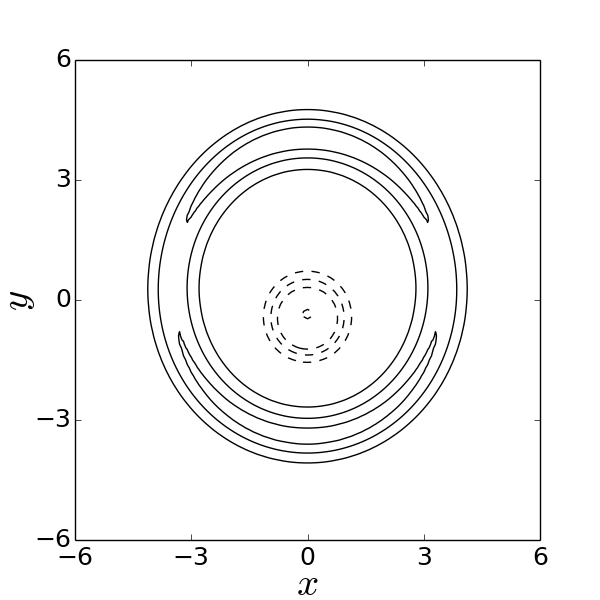}
\caption{Contour plot for the topological density $\skyrmion$ of the static skyrmionium (left) 
and the propagating skyrmionium with velocity $v=0.07$ (right). 
The contour levels plotted are chosen arbitrarily. Solid lines mean $\skyrmion > 0$ and
dashed lines $\skyrmion < 0$.}
\label{fig:skyrmionium_vel_topo}
\end{center}
\end{figure}

A further representation of the skyrmionium is given via the topological density in Fig.~\ref{fig:skyrmionium_vel_topo}.
The negative lump of topological density $\skyrmion$ moves off the center of the configuration for $v > 0$.
This gives manifestly a nonzero value $P > 0$ for the $x$ component of the linear momentum \eqref{eq:linear_momentum}.
For larger velocities the large axis of the elliptically shaped contours increases
and apparently diverges to infinity for $v \to v_c$.
We were able to numerically calculate the propagating configurations up to $v \approx 0.102$.

\begin{figure}[t]
\begin{center}
\includegraphics[width=4.2truecm]{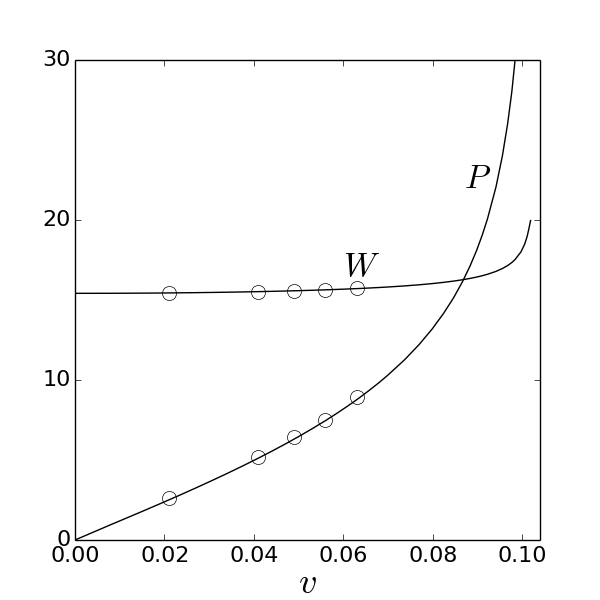}
\includegraphics[width=4.2truecm]{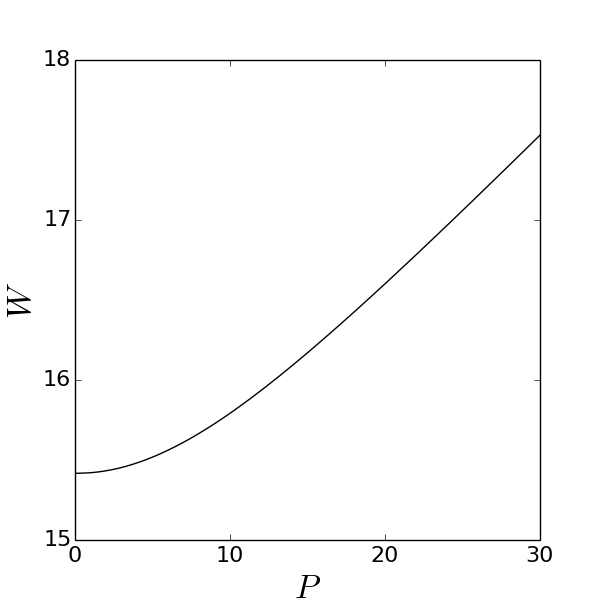}
\caption{(Left) Solid lines show the energy $\Energy$ of Eq.~\eqref{eq:energy} 
and linear momentum $P$ of Eq.~\eqref{eq:linear_momentum} for a steady-state propagating skyrmionium 
as a function of its velocity $v$.
The open circles show results from a direct simulation discussed in the text.
(Right) The energy $\Energy$ versus momentum $P$ for a steady-state propagating skyrmionium.
The curve is parabolic for low momenta in accordance with Eq.~\eqref{eq:energy_momentum_lowP}
but linear for large momenta in agreement with Eq.~\eqref{eq:linear_dispersion}.}
\label{fig:EandP}
\end{center}
\end{figure}

The energy $\Energy$ of a propagating skyrmionium can be calculated from Eq.~\eqref{eq:energy}
and its linear momentum $\bm{P}=(P,0)$ is given by Eq.~\eqref{eq:linear_momentum}. 
Fig.~\ref{fig:EandP} shows that both $\Energy$ and $P$ increase with velocity $v$.
For the static skyrmionium the energy $\Energy$ has a nonzero value while $P=0$.
Both quantities diverge to infinity as $v \to v_c$.
We could verify numerically that the group velocity relation
\begin{equation}  \label{eq:group_velocity}
\frac{d\Energy}{d P} = v
\end{equation}
holds to a very good accuracy ($\sim 1\%$). 

It is particularly interesting that $P$ is linear for values of the velocity $v \lesssim 0.06$,
and this motivates the definition of a mass $\mass$ for the skyrmionium from 
the Galilean relation
\begin{equation}
P=\mass v
\end{equation}
with the value $\mass = 117$ extracted from the numerical data.
Interestingly, the energy at low velocity is given by
\begin{equation}  \label{eq:newtonian_energy}
\Energy = \Energy_0 + \frac{1}{2}\mass v^2
\end{equation}
with the {\it same} effective mass $\mass$,
in agreement with the group velocity relation \eqref{eq:group_velocity}
which may also be written as $d\Energy/dv= v\, dP/dv$.
The corresponding energy-momentum relation reads
\begin{equation}  \label{eq:energy_momentum_lowP}
\Energy = \Energy_0 + \frac{P^2}{2\mass}
\end{equation}
at low momenta.
Fig.~\ref{fig:EandP} shows the numerically calculated energy-momentum relation and 
confirms the parabolic relation \eqref{eq:energy_momentum_lowP} at low momenta.
In the opposite limit $v \to v_c$ the energy-momentum relation is linear
\begin{equation}  \label{eq:linear_dispersion}
\Energy \approx v_c P
\end{equation}
which is also consistent with the group velocity relation \eqref{eq:group_velocity}.
Similar results have been recently obtained for a $\Skyrmion=0$ precessing magnetic droplet  
in a model without DM interaction \cite{BookmanHoefer_PRSA2014}.

In order to study the dynamics in the presence of an applied field
we use Eq.~\eqref{eq:linear_momentum} for the definition of the linear momentum
and Eq.~\eqref{eq:momentum_dynamics} applied for a field gradient \eqref{eq:field_gradient}.
We obtain
\begin{equation}  \label{eq:momentum_derivative}
\dot{P}_x = -\int \p_x \hext  (1-m_3)\, dx dy,\quad \dot{P}_y = 0
\end{equation}
which is the analogue of Newton's law for the case of a non-topological ($\Skyrmion=0$) skyrmion.
Under the field gradient the skyrmionium acquires a linear momentum along the gradient direction ($x$ axis)
according to Eq.~\eqref{eq:momentum_derivative} and confirms the Newtonian character of its dynamics.
For a pure gradient $\hext = gx$ we have
\begin{equation}  \label{eq:momentum_derivative_pure_gradient}
\dot{P}_x= -g\tmagn,\qquad \dot{P}_y = 0
\end{equation}
where $\tmagn=\int (1-m_3)\,dxdy$ is now the total magnetization of the skyrmionium.
For the static skyrmionium of Fig.~\ref{fig:skyrmionium_vecs} we have $\tmagn=73$
and thus $\dot{P}_x = -73 g$.

Finally, we have conducted a series of numerical simulations solving the initial value problem
for the conservative Landau-Lifshitz equation \eqref{eq:LL} using as initial condition the static
skyrmionium for $\anisotropy=3$.
We apply a field gradient \eqref{eq:field_gradient} with $g=-0.001,\; a=10$
and find that the $\Skyrmion=0$ skyrmionium indeed propagates along the direction of the gradient ($x$ axis)
in sharp contrast to the $\Skyrmion=1$ skyrmion propagation along the $y$-axis
shown in Fig.~\ref{fig:Q1_g0100_trajectory}.

\begin{figure}[t]
\begin{center}
\includegraphics[width=6.0truecm]{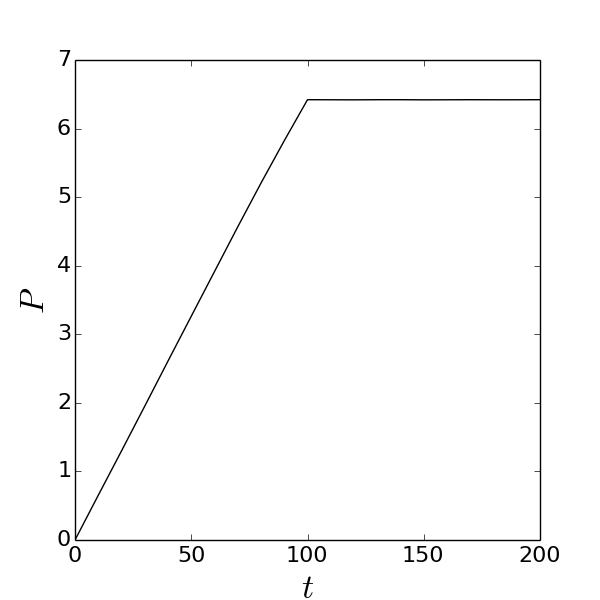}
\caption{Linear momentum $P$ as a function of time for a simulation
where a field gradient \eqref{eq:field_gradient} with $g=-0.001,\; a=10$ is applied to an initially static skyrmionium.
The field is applied for $0 \leq t \leq 100$, and is then suddenly switched off.
}
\label{fig:Q0_gradient_momentum}
\end{center}
\end{figure}

The field gradient is applied for the time interval $0 \leq t < 100$ and is then switched off.
Fig.~\ref{fig:Q0_gradient_momentum} shows the linear momentum as a function of time for the simulation.
The increase is linear while the field is applied, in accordance with the prediction of 
Eq.~\eqref{eq:momentum_derivative} or \eqref{eq:momentum_derivative_pure_gradient}.
From the numerical data we obtain $\dot{P}_x = 0.065$ which remains almost constant throughout the simulation.
This is in excellent agreement with the prediction obtained from Eq.~\eqref{eq:momentum_derivative}
 (deviation less that $1\%$).
The prediction of the simplified Eq.~\eqref{eq:momentum_derivative_pure_gradient} is $\dot{R}_x=0.073$
and shows a deviation from the simulation result due to the approximation of $\hext$ with a pure gradient.

\begin{figure}[t]
\begin{center}
\includegraphics[width=4.2truecm]{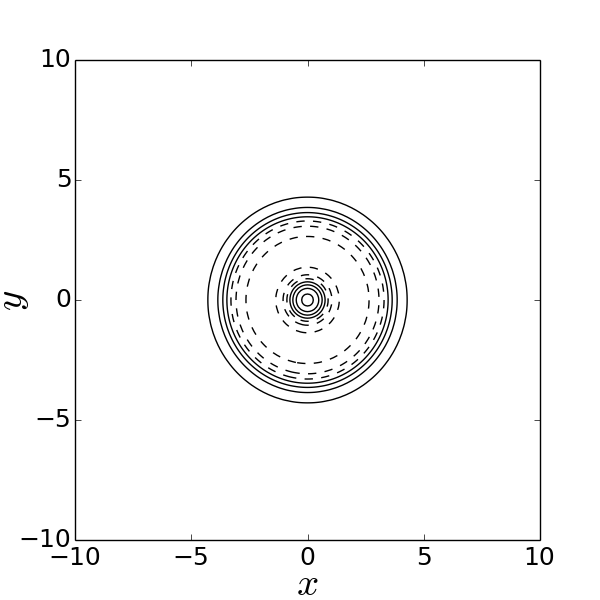}
\includegraphics[width=4.2truecm]{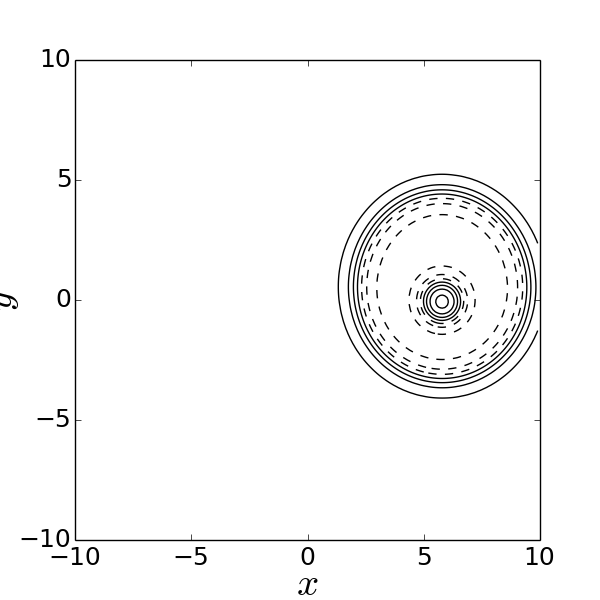}
\caption{Contour plots for $m_3$ for a skyrmionium under 
the field-gradient \eqref{eq:field_gradient} with $g=-0.001,\, a=10$.
(Left) The static $\Skyrmion=0$ skyrmionium of Fig.~\ref{fig:skyrmionium_vecs} is placed at the origin at initial time $t=0$.
(Right) The skyrmionium at simulation time $t=160$.
It propagates along the $x$ axis in the direction of the field gradient.
The contour levels plotted are $m_3=0.9,0.6,0.3,0.0$ (solid lines) and $m_3=-0.3,-0.6,-0.9$ (dashed lines).
}
\label{fig:Q0_gradient_m3}
\end{center}
\end{figure}

Fig.~\ref{fig:Q0_gradient_m3} shows a contour plot for the magnetic configuration at the initial time $t=0$ and at $t=160$.
We observe throughout the simulation that the configuration moves in a rather coherent way.
It has evolved from the initial static skyrmionium 
(shown in Figs.~\ref{fig:skyrmionium_vecs} and \ref{fig:skyrmionium_vel_m3} (left))
to a propagating one very similar to the those we calculated earlier in this subsection
(shown in Fig.~\ref{fig:skyrmionium_vel_m3} (right) for $v=0.07$).

At time $t=100$ we suddenly switch off the applied field and observe a free motion of the skyrmionium.
Its momentum remains constant and nonzero as shown in Fig.~\ref{fig:Q0_gradient_momentum} for $t>100$.
The skyrmionium propagates at a constant velocity confirming the Newtonian character of its dynamics.
We switch off the field for various times and measure the energy, linear momentum and velocity
of the freely moving skyrmionium.
The results are shown by open circles which have been superimposed on the plots for
the energy vs velocity and linear momentum vs velocity for steady-states in Fig.~\ref{fig:EandP}.
They almost coincide with the data calculated for steady-state configurations.

\section{Conclusions}
\label{sec:conclusions}

We have given a theoretical description of the dynamics of topological and non-topological magnetic solitons
in the presence of the Dzyaloshinskii-Moriya interaction.
When a field-gradient is applied
the $\Skyrmion=1$ skyrmion undergoes Hall motion perpendicular to the direction of the field gradient 
with a drift velocity proportional to the gradient, according to Eq.~\eqref{eq:skyrmion_velocity0}.
When the applied field is switched off the guiding center ceases to move while the skyrmion
undergoes cyclotron motion around its pinned guiding center. 
In contrast, the $\Skyrmion=0$ skyrmionium is accelerated in the direction of the gradient,
with acceleration proportional to the applied field (see Eq.~\eqref{eq:momentum_derivative}).
When the applied field is switched off the skyrmionium continues traveling with constant velocity $v$
in accordance with Newtonian dynamics of ordinary particles.

The dramatically different dynamical behavior between topological ($\Skyrmion \neq 0$)
and non-topological ($\Skyrmion=0$) solitons had been theoretically anticipated
\cite{Thiele_PRL1973,PapanicolaouTomaras_NPB1991,KomineasPapanicolaou_PhysD1996}
but a full demonstration had not been given
due to the lack of a system where both kinds of solitons could be conveniently studied. 
The DM materials support robust topological and non-topological magnetic solitons 
and thus appear to be particularly suitable for dynamical studies 
The dynamics can be induced by external fields, as in the present paper, 
or by magnon-skyrmion interactions 
\cite{Schuttegarst_PRB2014,IwasakiBeekman_PRB2014}.

The simulations presented show coherent propagation of solitons.
However, significant distortions of the magnetic configurations set in for larger field gradient.
For the simulation of the $\Skyrmion=0$ skyrmionium
we had to use a gradient as large as $|g|=0.1$ in order to observe significant distortions.

We have also given a calculation of skyrmionium in steady-state propagation with velocities
up to a maximum critical velocity: $v < v_c$.
The energy-momentum relation resembles that of a non-relativistic particle for low momenta 
(see Eq.~\eqref{eq:energy_momentum_lowP}) 
but becomes relativistic-like (linear) for large mometa (see Eq.~\eqref{eq:linear_dispersion}).

The picture derived in this paper with theoretical tools together with straightforward numerical calculations
is similar to the observed dynamics of magnetic bubbles \cite{MalozemoffSlonczewski}.
Topological bubbles ($\Skyrmion \neq 0$) are notorious for their skew deflection
in the presence of an applied magnetic field gradient.
The deflection is in a direction almost perpendicular to the field gradient,
in general agreement to the $90^\circ$ deflection of the $\Skyrmion=1$ skyrmion calculated in the present paper.
The deviation from the $90^\circ$ deflection is due to the presence of dissipation in bubble materials
and should also be anticipated for DM skyrmions.
Magnetic bubbles, stabilized by the magnetostatic interaction,
are easily deformable under external probes while the long-range nature of the interaction 
makes numerical simulations more complicated.
In contrast, chiral skyrmions are stabilized by the local DM interaction, 
thus they are robust and relatively easy to calculate numerically.
The relation between chiral skyrmions and bubbles have been discussed in 
Refs.~\cite{KiselevBogdanov_JPD2011,KiselevBogdanov_PRL2011,Ezawa_PRL2010}, 
where the differences in the energetics, in stability and of their cores were stressed.

\end{document}